\documentclass[sigconf]{acmart}

\usepackage{listings}
\usepackage[inline]{enumitem}
\definecolor{codegreen}{rgb}{0,0.6,0}
\definecolor{codegray}{rgb}{0.5,0.5,0.5}
\definecolor{codered}{rgb}{0.6,0,0}
\definecolor{backcolour}{rgb}{0.95,0.95,0.95}
\lstdefinestyle{mystyle}{
    backgroundcolor=\color{backcolour},   
    commentstyle=\color{codegreen},
    keywordstyle=\color{magenta},
    numberstyle=\tiny\color{codegray},
    stringstyle=\color{codered},
    basicstyle=\ttfamily\scriptsize,
    breakatwhitespace=false,         
    breaklines=true,                 
    captionpos=b,                    
    keepspaces=true,                 
    numbers=none,                    
    numbersep=5pt,                  
    showspaces=false,                
    showstringspaces=false,
    showtabs=false,                  
    tabsize=2,
    xleftmargin=0pt,
    xrightmargin=0pt,
    columns=fullflexible,
    aboveskip=0pt,
    belowskip=0pt,
    breakindent=2em,
}
\usepackage{array}

\usepackage{tikz}

\AtBeginDocument{%
  }

\copyrightyear{2026}
\acmYear{2026}
\setcopyright{cc}
\setcctype{by}
\acmConference[WTMC '26]{11th Workshop on Traffic Measurements for Cybersecurity}{November 15--19, 2026}{The Hague, Netherlands}
\acmBooktitle{11th Workshop on Traffic Measurements for Cybersecurity (WTMC '26), November 15--19, 2026, The Hague, Netherlands}
\acmDOI{10.1145/3846375.3849098}
\acmISBN{979-8-4007-3017-7/2026/11}

\usepackage[noabbrev,capitalise]{cleveref}
\usepackage[unicode]{hyperref}
\usepackage{subcaption}

\begin{document}

\title{How It's Made: Uncovering Detection Engineering Processes for Network Intrusion Detection Rules}

\author{Koen T. W. Teuwen}
\orcid{0000-0002-6490-4768}
\affiliation{%
  \institution{Eindhoven University of Technology}
  \city{Eindhoven}
  \country{The Netherlands}
}
\email{k.t.w.teuwen@tue.nl}

\author{Emmanuele Zambon}
\orcid{0000-0002-8079-4087}
\affiliation{%
  \institution{Eindhoven University of Technology}
  \city{Eindhoven}
  \country{The Netherlands}
}
\email{e.zambon@tue.nl}

\author{Luca Allodi}
\orcid{0000-0003-1600-0868}
\affiliation{%
  \institution{Eindhoven University of Technology}
  \city{Eindhoven}
  \country{The Netherlands}
}
\email{l.allodi@tue.nl}

\begin{abstract}
Many Security Operations Centers rely on signature-based Network Intrusion Detection Systems like Suricata, yet detection rule engineering remains understudied. We investigate this process by introducing SuriCap, a platform for rule engineering exercises, and hosting CTF-style workshops where $60$ participants, trained MSc students, and experienced SOC professionals, created rules for four scenarios. Participants produced $3146$ valid rules, enabling analysis of their methods, performance, and iteration patterns. Surprisingly, prior experience had limited impact on rule quality, suggesting that less experienced engineers can produce rules comparable to experts. We also observed challenges in generalizing rules beyond available tests, underscoring the need for sufficient labeled data. From our study, we identify three phases and a common pattern in rule engineering, offering SOC managers insights to improve their processes and expectations of engineer expertise.
\end{abstract}



\maketitle

\section{Introduction}
\label{sec:introduction}
Organizations constantly face security threats and take security measures to mitigate their effects.
Among these measures, Intrusion Detection Systems (IDSs) such as Suricata~\cite{Suricata, alert-alchemy, ruling-the-rules, ruling-the-unruly} are commonly deployed to monitor network traffic traversing relevant network links.
Modern Security Operations Centers (SOCs) face challenges at both organizational and technical levels~\cite{matches-and-mismatched-socs, vielberth20-soc-systematic-study-open-challenges}. Among these, the ingestion of new rules and updating of old ones is a key concern as assessing rule quality and relevance is hard, and testing is expensive or impossible. Partially addressing these problems, some SOCs engineer their own in-house detection rules~\cite{alert-alchemy}. Widely available commercially engineered rulesets are sold as a product~\cite{et-pro-rules,snort-products}, and open community rulesets exist as well~\cite{et-open-forum-signatures}, both of which can be used in conjunction with in-house rulesets to increase coverage~\cite{alert-alchemy}.
Although industrial rule engineering processes lack extensive documentation, limited available public evidence suggests that rules are usually engineered given concrete samples of malicious and benign behavior~\cite{panther-detection-engineer-rule-writing}.
Previous work has already investigated high False Positive (FP) rates within SOCs~\cite{99-false-positives}, in relation to rulesets~\cite{ruling-the-rules}, and to rules~\cite{ruling-the-unruly}. 
Although rule performance primarily depends on how a rule is developed, the current literature does not explore how rule engineers initially develop rules and how the adopted engineering strategies impact rule performance. Similarly, no study reports on the role of experience in rule quality or whether experts engineer rules differently from novices.
Recent work covers the design aspects of detection rules~\cite{ruling-the-unruly}, but to our knowledge, no studies evaluate the rule creation process.

\looseness=-1
We address these gaps by studying how network intrusion detection rules are engineered from incidents. Inspired by work in related fields~\cite{decomperson, remind, humans-vs-machines}, we organize several workshops akin to Capture the Flag (CTF) competitions in which participants engineer Suricata rules. To provide a realistic setup, we collaborate with a commercial SOC that provides expertise and training material used for their own analysts. Study subjects are recruited from a cohort of students at different levels, BSc computer science students with limited security training and MSc students with formal security training including intrusion detection, as well as expert engineers employed at the collaborating SOC. Participants access \textsc{SuriCap}, a platform developed for this study, where they engineer rules for different attack scenarios, submit them to receive feedback on correctness and performance, and iteratively refine them. We collect data on each participant's progress, yielding a rich dataset describing the engineering process, how rules evolve, and what strategies participants employ.
Concretely, this study answers the following research questions:
\begin{itemize}[leftmargin=2.5em]
    \item[RQ1] How are network intrusion detection rules engineered?
    \item[RQ2] How does the experience of rule engineers affect the engineering and performance of the developed rules?
\end{itemize}

\section{Background and Related Work}
\label{sec:background}

\paragraph{Security Operations \& Rule-based Intrusion Detection}
\label{sec:background_soc}
\label{sec:rulesets}
A \textit{Security Operations Center (SOC)} is tasked with monitoring to detect and resolve security breaches~\cite{vielberth20-soc-systematic-study-open-challenges}.
SOCs utilize network or host monitoring tools to collect data on assets, employing \textit{Host-based Intrusion Detection Systems (HIDS)} to detect malicious activity on endpoints, and \textit{Network-based Intrusion Detection Systems (NIDS)}, the focus of this study, to detect malicious network activity to/from these devices.
Suricata~\cite{Suricata-docs}, which is the focus of this work, is the de facto standard
NIDS and uses \textit{rulesets} containing \textit{signatures} for known threats.
Snort~\cite{snort} is another rule-based NIDS that uses a highly similar rule syntax for historical reasons.
Each Suricata rule begins with a \textit{header} specifying the \textit{action}, \textit{protocol} and \textit{direction} of inspected packets, followed by a \textit{body} containing additional options. Crucially, \textit{detection options}  specify which buffer is matched with which string, bytes, or regular expression. 
Some buffers support inspection after decompression or decoding, without requiring manual field location.
\textit{Content modifiers} change how content is matched (e.g., case sensitivity). Rules may combine positive and negative matches or specify a \textit{distance} between them.
The \textit{flow} keyword restricts inspection to packets sent by servers or clients, while keywords like \textit{flowbits} and \textit{xbits} enable stateful detection across packets or connections. Rules can also use \textit{threshold}, e.g., to control triggering frequency, and non-detection keywords like \textit{metadata} to aid alert interpretation.
Rule engineers may use arbitrary combinations of these features, where keyword order matters only for some detection options.
Rules are used to detect malicious behavior via, e.g., Indicators of Compromise, bytes indicative of known exploits, or protocol deviations~\cite{et-category-descriptions}.

Prior literature studying rule(set) evolution~\cite{ruling-the-rules} suggests a minority of rules causes the majority of alerts, and that most rules receive updates only to metadata rather than detection options. 
Since most rules never receive updates, the initial engineering process is critical, motivating our focus on this process before rules are first released.
Related research on SOC workflows~\cite{alert-alchemy} found that many SOCs dedicate significant resources to engineering in-house rules tailored to client environments. Unlike in-house rulesets, open and commercial rulesets are not designed to operate in a specific environment. Open rulesets~\cite{et-open-forum-signatures} may include rules submitted by engineers with varying experience~\cite{bitsight-hunting-privateloader}.
SOCs commonly tune external rulesets by disabling rules~\cite{alert-alchemy} while re-engineering loops from reported FPs are uncommon, making the initial rule engineering process all the more influential on rule performance.
Other work describes a SOC analyst attributing FPs to poor rule engineering~\cite{turning-contradictions}.
The Suricata community~\cite{suricata-style-guide, snort-suricata-guide} has developed quality guidelines focusing primarily on syntax and structure. A recent study proposed six design principles to increase rule specificity while maintaining coverage~\cite{ruling-the-unruly}, addressing features like alert throttling, matching content and method generalizability. However, these works address what rules should look like rather than how to derive such rules or how to apply the principles when engineering new rules.

Despite the recognized importance of detection engineering~\cite{splunk-detection-engineering,truesec-detection-engineering}, little is known about the initial rule engineering process. 
Prior work suggests using unit tests with positive and negative cases, backtesting on historical benign data, and staged deployment to prevent FPs~\cite{panther-detection-engineer-rule-writing}.
To the best of our knowledge, our work presents the first academic study of the rule engineering process.

\paragraph{CTF-style Activities for Research}
\label{sec:jeopardy}
\looseness=-1
Research methods similar to ours have previously been applied to study reverse engineering and malware classification, where Jeopardy-style workshops are used to study how people address these tasks. One such work studies  `perfect decompilation' whereby participants work towards a predefined solution (i.e., pre-compilation source code)~\cite{decomperson}.
Another study on reverse engineering investigates the effects of experience on strategies and performance~\cite{remind}. Unlike \cite{decomperson}, little feedback is given to participants and the platform uses a restricted focus viewer technique, blurring parts of the code to measure attention, across three challenges.
A further related work studied how people classify malware compared to machines~\cite{humans-vs-machines}, investigating which features people inspect to derive conclusions about file maliciousness.
Although the related work differs from our detection engineering domain (given the lack of a perfect solution), we draw inspiration from these studies, particularly the use frequent submissions to better capture the overall process. Stimulating this, we incorporate syntax highlighting, clear feedback on errors, and a `baby' challenge to help users get acquainted with the platform. Differentiating \textsc{SuriCap} and uniquely relevant to intrusion detection is its ability to assess the generalizability of submissions beyond the tests visible to participants.

\section{Methodology}
\label{sec:experiment_design}
\begin{figure}[htbp]
    \includegraphics[width=1.0\linewidth]{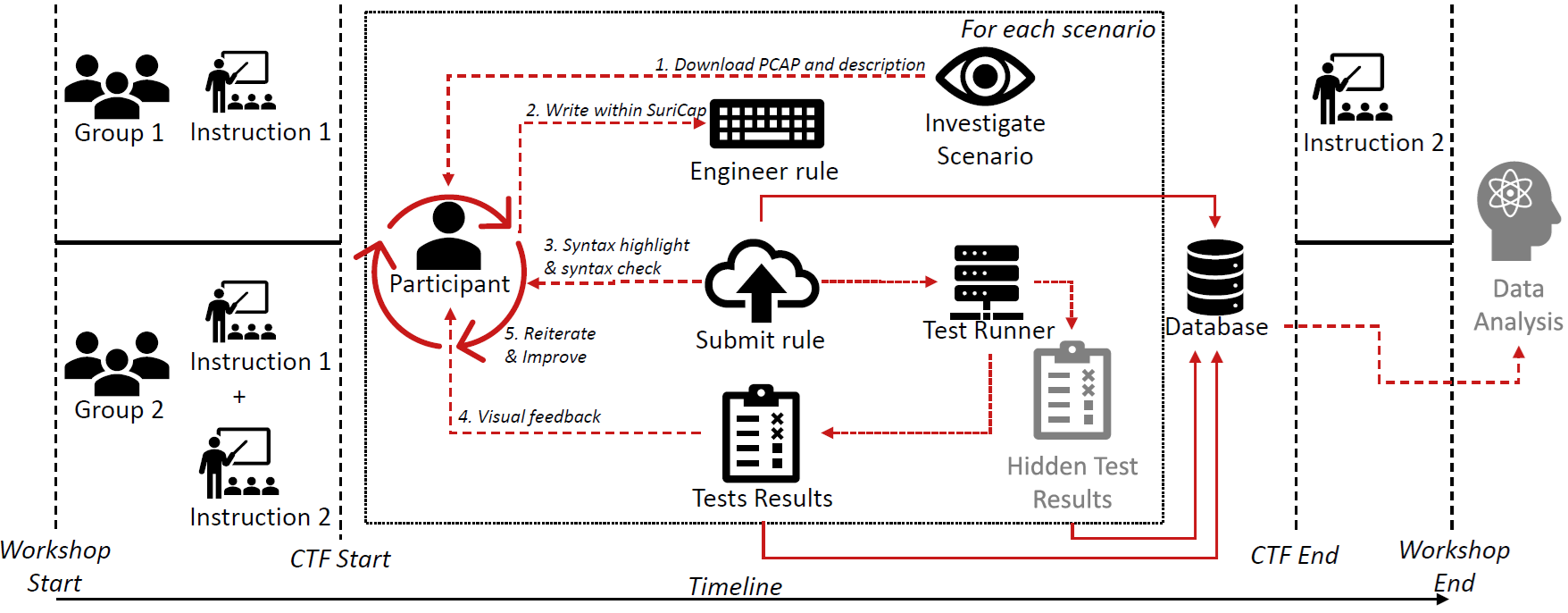}
    \centering
    \caption{Experiment design to support answering of RQs.}
    \label{fig:methodology}
    \Description{Overview of the study workflow: participants receive training, inspect attack traffic, engineer and submit Suricata rules through SuriCap, receive feedback from visible tests, and are evaluated on hidden tests. The study compares participants with different experience levels and compares a design-principles treatment group with a control group.}
\end{figure}

To answer the RQs presented in \cref{sec:introduction}, we devise a methodology, summarized in \cref{fig:methodology}, supported by a dedicated measurement platform that allows us to capture data at-scale.
The platform is designed to be non-intrusive, allowing participants to exhibit natural behavior, rather than forcing specific actions.
Inspired by CTFs in related fields~\cite{decomperson}, we organize the activity as a CTF where participants are presented with various challenges in the form of \textit{scenarios} and compete with each other within the time of the activity for the best coverage and specificity of their engineered rules.
Participants tackle scenarios by inspecting network traffic packet captures (PCAPs) and engineering Suricata rules to detect those scenarios. During this process, participants receive feedback on their submissions from a limited number of \textit{visible tests} which they can use to refine their rules.
Additional \textit{hidden tests} are performed to reflect real-world scenarios~\cite{panther-detection-engineer-rule-writing}. We include both positive (malicious) and negative (benign) (in)visible tests to assess coverage and specificity, which are commonly regarded the most important performance dimensions of detection rules~\cite{ruling-the-rules, alert-alchemy, ruling-the-unruly}.
This mimics the unit and backtesting~\cite{panther-detection-engineer-rule-writing} described in \cref{sec:rulesets} whereby rules are initially engineered given a positive and a negative test, and then further tested on unseen data in staging and production environments.

\looseness=-1
Participants physically attend the two hour workshop to address challenges, while a leaderboard displays current scores on the visible tests. To make the workshops educational whilst not biasing experimental outcomes, participants can, at any time during the workshop, ask the organizer of the CTF to clarify objectives or for assistance with the Suricata rule language, but will receive no directions for solutions to challenges. Before workshops, participants receive personalized instruction through a video lecture and a handout.

\looseness=-1
The experiment is run using a gamified platform, \textsc{SuriCap} to collect data on rules while they are being engineered enabling research on these processes from start to end (RQ1).
From the participants' point of view, \textsc{SuriCap} enables competition whilst easing rule development through automated feedback. The workshop is freely and voluntarily accessible to everyone. We advertise the workshop in cybersecurity courses delivered by European technical universities. To access the workshop, interested participants complete an intake questionnaire to gauge relevant background and experience (RQ2).

We randomly divide participants in two groups to evaluate if knowledge of rule design principles~\cite{ruling-the-unruly} impacts rule engineering (outcomes).
A randomly defined treatment group receives an additional instruction on the design principles prior to the activity, whereas the control group only receives the standard SOC training.

During the activity, participants are  asked to engineer Suricata rules to detect malicious traffic by raising a single alert while remaining insensitive to benign traffic.
Participants know the attack technique (e.g., Command \& Control over encrypted channel) contained in each scenario, but are not given any specifics on attack triggers or characteristics they could use for detection. Participants can inspect a PCAP to determine suitable detection characteristics for engineering rules.
Participants can submit the rules they are developing to the backend system to receive automated feedback in the form of success/failure to (not) detect malicious/benign traffic in the visible tests, which they can use to to further refine their rules. Additional information such as error messages or suricata CLI output are also provided to the user through the interface.
Participants could submit rules as often as desired.
The hidden  tests are run to measure the coverage and specificity beyond the PCAPs visible to the participants. This replicates real-life circumstances of the rules being deployed against data they were not tested against.
To reproduce this in full, the visible test tests rules against the oldest samples, whereas the older, often more complex malicious behavior  (e.g., using evasion techniques) is tested against by the hidden test.
Each test has associated Suricata configuration options to ensure variables that can be used in rules (e.g., \texttt{\$HOME\_NET}) are defined appropriately for the sensor setup and scenario represented by the test.
Hereafter, 
we first describe ethical considerations, how we recruit participants and the process they undergo for the experiment. Subsequently, 
we detail scenarios for which we measure the rule engineering process and elaborate on the measurement platform.

\paragraph{Ethics considerations}
This research was carried out with approval from our institution’s ethical review board under the approval number ERB2024MCS34. We obtained explicit informed consent from all subjects whose data are used in this study and assured that the research would not affect their study/work conditions. Participation in the activity is non-obligatory and solely encouraged through the educational and entertaining aspects. Moreover, participants of the activity are not obliged to participate in the research.
The experiment design, including the presence of hidden tests, is explained prior to the activity, and sample solutions are provided afterward. Measurement of user input occurs non-intrusively through explicit user input. After data collection and correlation, personal information is removed so that researchers cannot use the data to identify individuals to whom data relates.

\paragraph{Subjects Recruitment}
\looseness=-1
We recruit participants through academics at our own and other institutions, who helped us organize the workshop locally and announce the workshop to potentially interested students. We also reached out to student communities (i.e., CTF associations) at those universities to further promote the workshop. An announcement was also shared within their respective communities.
Although we aimed primarily at recruiting MSc students following a cybersecurity program, the workshop was explicitly open to anyone, resulting in the recruitment by word of mouth of also the BSc students and PhD graduates.
The recruited participants fit the job profile for junior SOC roles~\cite{kathryn-22} and are similar to rule engineers who occasionally contribute to open rulesets.
This led to the organization of five workshops at four institutions\footnote{\scriptsize Extraordinary circumstances resulted in the moving of one workshop, which negatively affected participant turnout, which prompted us to organize a second workshop at the same institution.}, in which a total of $60$ people with a wide variety of backgrounds participated. $85\%$ had previously completed a BSc degree or equivalent as highest education level.
$62\%$ of the participants learned about the activity through a cybersecurity MSc course they attended, whereas $30\%$ learnt about the activity through a group of CTF enthusiasts.
These statistics suggest that the majority of all participants are indeed currently following a MSc program with a focus on cybersecurity. $80\%$ of the participants identify as male, whereas $17\%$ identify as female.

Further, three experts from the collaborating SOC accepted to join the experiment.\footnote{\scriptsize While recruitment of senior analysts is scarce due to limited availability, the experts serve as a meaningful separate comparison benchmark against other participants.}
The experts all have years of experience in incident detection and detection engineering using Suricata.
One expert has 6 years of experience and is currently a senior analyst at the collaborating SOC. Another expert has 3 years of SOC experience related to triage and infrastructure. The third expert has 5 years of experience running the SOC's detection infrastructure.

\paragraph{Preparatory Instructions}
\looseness=-1
One week before the experiment, participants receive a preparatory video lecture. The video lecture is based on the training material provided by the collaborating SOC for the training of their own analysts, and includes information on the experiment design, required tools, and the platform the participants will interact with.
The treatment group receives an additional instruction on the design principles from ~\cite{ruling-the-unruly}.
The instruction (Appendix~\ref{app:instruction}) familiarizes participants with relevant prerequisite knowledge on the tools they will be working with, including Suricata and the platform.
All participants receive a handout with the material covered during the video lecture, which is also made available on the platform during the exercise. The usage of a video lecture rather than an in-person lecture ensures that all participant in the control/treatment group receive identical instructions.

\paragraph{User questionnaire}
In order to control for possible confounding factors in our subject pool we design a questionnaire to gauge relevant prior experiences of participants that may affect rule engineering processes or experimental outcomes. The questionnaire (Appendix~\ref{app:questionnaires}) covers network protocols (e.g., TCP, TLS), and offensive computer security, to the extent that these relate to the scenarios presented in \cref{sec:scenarios}. These topics have previously been identified as relevant for Tier 1 SOC analysts~\cite{radutest}. The questionnaire also covers Wireshark, intrusion detection, and Suricata, which are the core technologies participants are expected to utilize during the activity.
The questions about offensive computer security specifically relate to the tactics of the scenarios discussed hereafter.
We devise questions to measure experience with relevant topics to avoid reliance on self-assessment. For example, instead of asking participants directly how experienced they are in working with network protocols, their experience is gauged using questions like: \textit{``how does TCP ensure delivery of packets in the correct order?''}
We also collect demographics like age and education level to describe the population and inquire how frequently participants attend CTFs. A control question was included to assess respondent reliability.

\paragraph{Scenarios design}

\label{sec:scenarios}
To ensure our analysis covers a wide range of NIDS rules, the experiment consists of one ``baby'', and three main scenarios, each focusing on one common MITRE ATT\&CK tactic~\cite{mitre-taxonomy} that can be detected by network monitoring. Furthermore, the scenarios cover rules for which detection rules are commonly engineered.\footnote{\scriptsize $63\%$ of all rules from the ETOPEN ruleset focus on the HTTP and TLS protocols.}
Concretely, the scenarios focus on Reconnaissance (i.e., scanning for unintended information disclosure), Initial Access (i.e., payload delivery through a download following malspam), and Command and Control (i.e., C2 communications from a RAT). These are chosen to be representative of tactics commonly detected by NIDSs~\cite{ruling-the-unruly}; further, similarly to \cite{remind}, we must find a balance between the time we expect participants to invest in the activity and the data we can collect.
Scenarios are designed to allow for both simple and complex solutions, which although may both pass the visible tests and appear successful from a participant's point of view, may perform differently on the hidden tests.
In designing the tests for scenarios, the working principle is that visible test data is chosen that would likely be available to a rule engineer when implementing a rule, which means it is limited~\cite{panther-detection-engineer-rule-writing}. Following this principle, the oldest network traffic, or traffic concerning the most prevalent system type are visible to participants whereas more recent data or traffic corresponding to more rare systems are hidden from participants. This ensures the realism of the data availability at the time of rule engineering, as the `training process' (i.e., the rule engineering) cannot rely on data from the future~\cite{arp-22}.

\looseness=-1
All participants start with a `Baby' warm-up scenario, followed by the three main scenarios in randomized order.
Once a user has obtained the maximum possible score on the visible tests the next scenario is unlocked. To prevent circumstances in which a user may be deterred because they fail to obtain this score, the next scenario is  unlocked after a user spent $5$ minutes working on that scenario.\footnote{\scriptsize This proved to be mostly irrelevant: in $89\%$ of cases participants only moved to the next scenario after passing the visible tests on their current scenario; conversely, participants who moved on to the next scenario without passing these tests spent on average $22$ minutes on that scenario.}
The user is free to go back to any previously unlocked scenario anytime.
Hereafter, we exemplify how the \textit{Information Disclosure} scenario and the benign tests (shared among all scenarios) are derived. The derivation of the other scenarios is presented in Appendix~\ref{app:additional_scenarios}.

\noindent\textit{Information Disclosure.}
The \textit{Information Disclosure} scenario covers an unintended information leak detailing web server configurations through the built-in \texttt{phpinfo} function, from which adversaries can derive installed software, as well as server configuration settings.
A SOC may choose to detect this type of behavior as precursor of follow-up scans or attacks, or for forensic reasons during incident reconstruction\footnote{\scriptsize For example, see ET rule $2019526$.}.
In collaboration with SOC experts, we collect PCAPs from servers with various PHP versions where disclosure was successful, with and without the use of evasive methods, like applying base64 encoding to the URL~\cite{xss-filter-evasion}.
Similarly, we collect PCAPs from servers where information disclosure failed to test for FPs.
We also collect PCAPs of access to websites with similar content but without information disclosure, such as pages documenting the \texttt{phpinfo} function, which could trigger FPs, in case the rules lack specificity in their match definitions.
Participants see the performance of the rules they submit evaluated only against the capture corresponding to a successful disclosure against the server running the oldest PHP version in line with our working principle. The remaining $23$ PCAPs, which represent more recent or more complex instances of the same behavior, are used as hidden tests.

To exemplify how the various tests behave with respect to different rules, we explain how two examples (see \cref{fig:phpinfo-rules}) may fail to trigger true positives or trigger FPs.
The `good' example
obtained a perfect F1-score of $1.0$ across the all tests combined whereas the `suboptimal' rule
obtained an F1-score of $0.56$. The primary difference is that the good rule detects server responses instead of client requests such that only successful attempts are detected, in line with the design principles from~\cite{ruling-the-unruly}. Instead, this rule also triggers on several tests corresponding to unsuccessful attempts, thus resulting in FPs. 
Moreover, the suboptimal rule only captures one of the possible ways in which the \texttt{phpinfo} function may be invoked, thus missing cases in which the URL was encoded, or the \texttt{phpinfo} function was invoked from web resources not named ``phpinfo''.

\begin{figure}[tbp]
\begin{minipage}{\columnwidth}
\begin{lstlisting}
alert http $HOME_NET any -> $EXTERNAL_NET any (
    msg:"Good phpinfo rule"; flow:established,to_client;
    file.data; content:"<title>"; content:"phpinfo"; within:20;
    threshold:type both, track by_src,count 1, seconds 180;)
\end{lstlisting}
\vspace{0.10in}
\begin{lstlisting}
alert http $EXTERNAL_NET  any -> $HOME_NET any (
    msg:"Suboptimal phpinfo rule";
    http.request_line; content:"/phpinfo."; flow:to_server;)
\end{lstlisting}
\end{minipage}
\caption{`Good' (top) and `Suboptimal` (bottom) rule detecting Information Disclosure}
\Description{Two Suricata rules for detecting information disclosure. The top rule uses established traffic to the client, matches title and phpinfo content in file data, and limits alerts with a threshold. The bottom rule only matches a phpinfo path in the request line and lacks specific content and alert-throttling conditions.}
\label{fig:phpinfo-rules}
\end{figure}

\noindent\textit{Benign Traffic.}
We include several PCAPs that contain only benign traffic from \cite{stratosphereips}.
The benign PCAPs are chosen to include a variety of HTTP traffic to test for FPs from any rules that may aim at detecting the HTTP traffic from the \textit{Information Disclosure} scenario and \textit{Malspam Download} scenarios. Similarly, benign PCAPs that contain TLS traffic are beneficial in evaluating rules for the \textit{RAT} scenario.
These PCAPs contain no malicious traffic, and high-quality rules are not expected to raise any FPs on them. In addition to the scenario PCAPs selected to be made available to participants, we also include a capture containing benign traffic originating from a Windows 7 PC.
This capture contains a variety of benign behaviors, like usage of a web browser, cloud storage providers to download files, and internet radio, along with traffic generated by background services running on the device.
In line with our working principle, we consider this capture to be representative of (older) common traffic, since Windows is the prevalent desktop operating system~\cite{statcounter-desktop-os-share}.

Three additional more recent benign PCAPs are selected as hidden tests to evaluate whether rules generalize to unseen benign traffic without raising FPs. Another more recent capture from a Windows 7 PC containing similar traffic, but also video streaming, is leveraged as an addition to the older capture to evaluate whether engineered rules may raise more FPs on traffic collected after the traffic for which the rule was originally engineered.
To assess whether rules may trigger FPs on other Operating Systems and/or previously unseen protocols, we select a capture containing normal traffic from a Debian host performing various normal actions including P2P Deluge downloads, web browsing, video streaming, and using Jabber.
Furthermore, to assess the potential for FPs when interacting with various web servers that potentially run different software with different configurations, we select a benign PCAP that contains interactions with Alexa top $1000$ domains.

\begin{figure*}[htbp]
    \centering
    \includegraphics[width=0.8\textwidth]{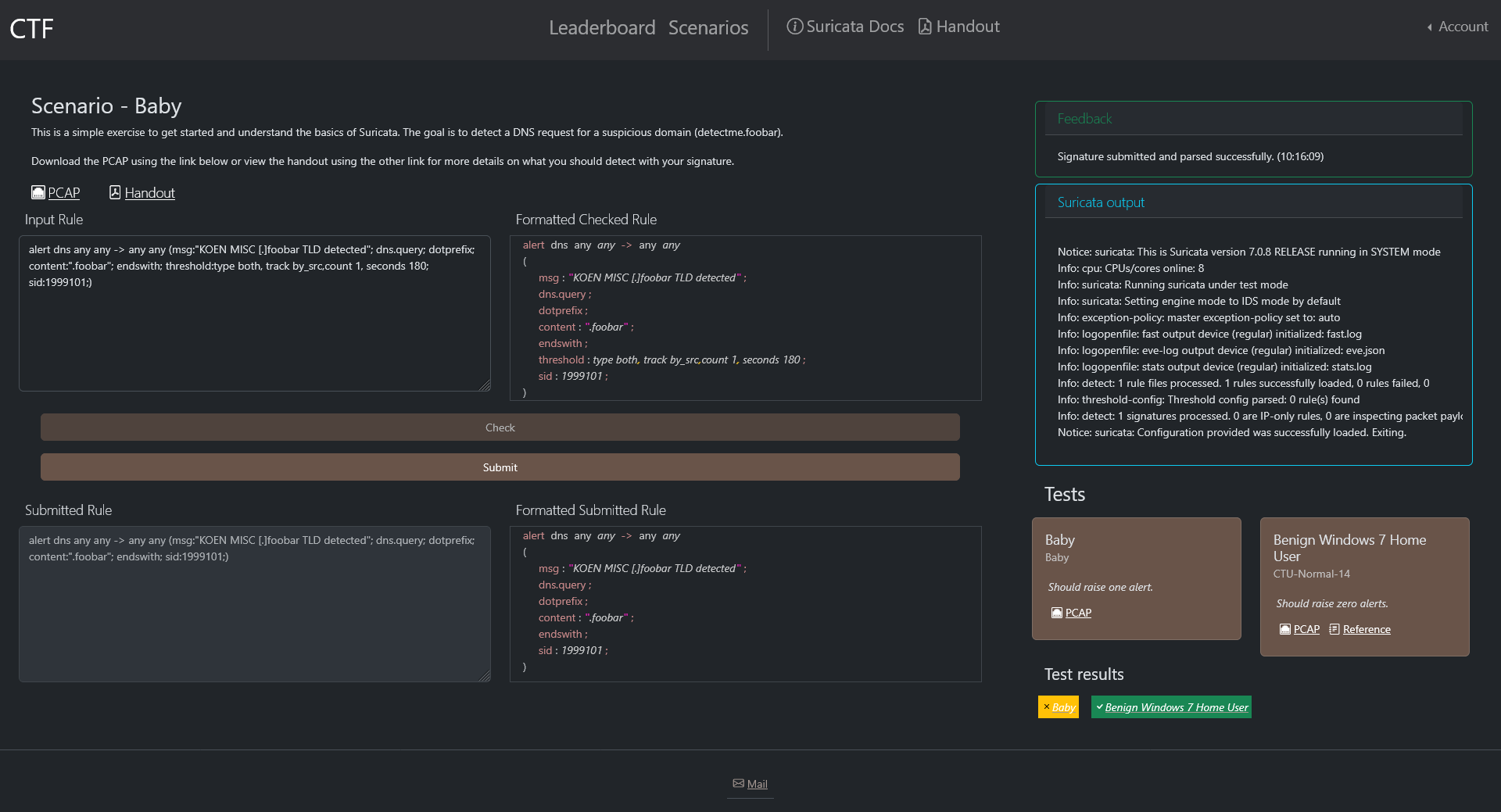}
    \caption{A screenshot of the UI of \textsc{SuriCap}, showing the menu bar containing resources such as the documentation and the handout, the scenario description, original and syntax highlighted user inputs, and test descriptions and results.}
    \label{fig:scenario-ui}
    \Description{Screenshot of the SuriCap interface. A menu bar provides documentation and handout links; a scenario description and rule editor appear on the left; syntax-highlighted rule input, Suricata feedback, and visible test descriptions and results appear on the right.}
\end{figure*}

\paragraph{The \textsc{SuriCap} Platform}
\label{sec:measurement_platform}
To execute the methodology, a measurement setup is required to observe and collect data on the rule engineering process.
To this end, we propose \textsc{SuriCap}: a measurement platform to study and evaluate Suricata rules.
\textsc{SuriCap} acts as a wrapper for Suricata: it captures user input in the form of rules and automatically provides feedback to the user while hiding Suricata's CLI and configuration.
\cref{fig:scenario-ui} provides an overview of the \textsc{SuriCap} interface.
The menu bar at the top provides easy access to the instruction handout and the Suricata documentation~\cite{Suricata-docs}.
On the top left is provided a description of the current scenario.
Each scenario has at least one associated PCAP containing malicious behavior that the rule developed by the participants should detect.
Scenarios can contain additional instructions in the form of a textual description or a file (i.e., PDF) to provide relevant information to users, such as the context from which the traffic was captured. All scenarios described in \cref{sec:scenarios} contain a brief overview, without suggesting specific characteristics for detection.
Users can download the network traffic from the User Interface (UI) to inspect the PCAPs using Wireshark. 
When a user \textit{checks} a rule written in the upper-left input box on the left, feedback in the form of syntax highlights in the upper right and Suricata output in the upper right indicating any potential syntactical errors.
When a user \textit{submits}, the lower input boxes on the left will display the most recently submitted rule, and tests will be run in the back-end to render test results in the far bottom right in addition to the feedback generated by a check. The tests and results thereon are shown on the right, including the expected test outcome and PCAPs.

Similarly to prior work~\cite{decomperson}, \textsc{SuriCap} provides swift feedback whenever a rule is submitted to incentivize frequent rule submission and highlights syntax by formatting rules, splitting options to separate lines and coloring keywords and special characters.
The platform also indicates whether Suricata was able to successfully load the rule, showing the user detailed Suricata output to correct invalid rules.
Users receive color-coded (i.e., green for successes, red for failures) feedback indicating whether the malicious behavior is correctly detected, and whether benign data triggered FPs. Additional details regarding the number and nature (i.e., Suricata Fast and EVE logs) of FPs can also be accessed through the UI. 

To reinforce the gamification element and similarly to~\cite{decomperson}, during the workshop participants are awarded a score based on the tests to compete with other participants on a leaderboard.
The score of participants is computed using their best submission instead of their most recent submission.
In case multiple participants obtain the same high-score, the participant who first obtained that score will win.
Participants are informed that at the end of the workshop the `hidden leaderboard' based on all tests including the hidden tests will be revealed to encourage them to continue working on their detection rules after they have passed all visible tests to enhance rule generalizability. The winners receive no awards beyond the fame resulting from the announcement to encourage intrinsically motivated participants.
Through the provision of these various types of feedback and the inclusion of time in the scoring method, we stimulate frequent submissions to establish a detailed dataset.

\section{Data Analysis}
\label{sec:results}
We organized $5$ CTF sessions to collect data from $60$ participants with varying experience levels (RQ2) who jointly submitted $3146$ valid rules. From these rules, we consolidate $5176$ individual changes (e.g., adding or modifying a Suricata option) that describe engineering processes followed by participants (RQ1). Combined with the $15751$ test results for all valid rules (i.e., rules parsable by Suricata), we evaluate the engineering process that analysts follow to develop their rules and experience effects in relation to rule quality.

\paragraph{Experiment progression}
\label{sec:experiment_progression}
\cref{tab:data_statistics} presents an overview of the data collected on rule engineering processes for each scenario and across all combined scenarios. We can see that most participants obtained good F1-scores on the Baby scenario and passed the visible tests on average after $7$ minutes. Some participants had to get acquainted with the Suricata syntax at the beginning of the activity, as depicted by the lower fraction of valid submitted rules for the Baby scenario.
Note that many participants failed to reach optimal F1-scores for the main scenarios and there is a high variance therein. The recall scores, which are lower than the precision scores, suggest that participants struggle more with coverage than specificity. These results suggest that different rules result in different specificity/coverage and that many rules could possibly be improved.
Furthermore, the data suggests that participants struggle the most with engineering performant rules for the Malspam Download scenario, as indicated by the lower F1-scores, more time required to pass the visible tests, and the increased rule submissions.
Note that recall scores obtained by participants ($0.59$) are only marginally lower than those obtained by the expert group ($0.65$) or relevant public rules~\footnote{\scriptsize ET rules $2035607$ and $2035595$ obtain recall of $0.67$ on the RAT scenario and ET rule $2019324$ obtains a recall of only $0.25$ on the Malspam Download scenario.}.

\begin{table*}[htpb]
    \caption{Statistics describing the engineering process of the average (mean and SD) participant.
    }
    \label{tab:data_statistics}
    \resizebox{\textwidth}{!}{
    \begin{tabular}{lrrrrrrr}
        \toprule
        Scenario & F1-score & Precision & Recall & \shortstack{Number of \\ Design Principle \\ Issues} & \shortstack{Time (scnds) \\ to Pass \\ Visible Tests} & \shortstack{Number of \\ Submitted Rules} & \shortstack{Fraction of \\ Valid \\ Submitted Rules} \\
        \midrule
        Baby & $0.98\pm0.13$ & $1\pm0$ & $0.98\pm0.13$ & $3\pm1$ & $430\pm592$ & $22\pm18$ & $0.66\pm0.19$ \\
        Information Disclosure & $0.59\pm0.23$ & $0.88\pm0.16$ & $0.51\pm0.25$ & $3\pm1$ & $667\pm685$ & $17\pm18$ & $0.73\pm0.22$ \\
        Malspam Download & $0.48\pm0.26$ & $0.96\pm0.16$ & $0.37\pm0.27$ & $3\pm1$ & $1273\pm972$ & $22\pm25$ & $0.74\pm0.22$ \\
        RAT & $0.60\pm0.28$ & $0.94\pm0.20$ & $0.52\pm0.27$ & $3\pm1$ & $968\pm689$ & $19\pm17$ & $0.78\pm0.18$ \\
        \midrule
        All scenarios & $0.67\pm0.15$ & $0.84\pm0.16$ & $0.59\pm0.19$ & $10\pm4$ & $2931\pm1336$ & $75\pm48$ & $0.66\pm0.19$ \\
        \bottomrule
    \end{tabular}
    }
    \Description{}
\end{table*}

\begin{figure*}[htbp]
    \includegraphics[width=1.0\textwidth]{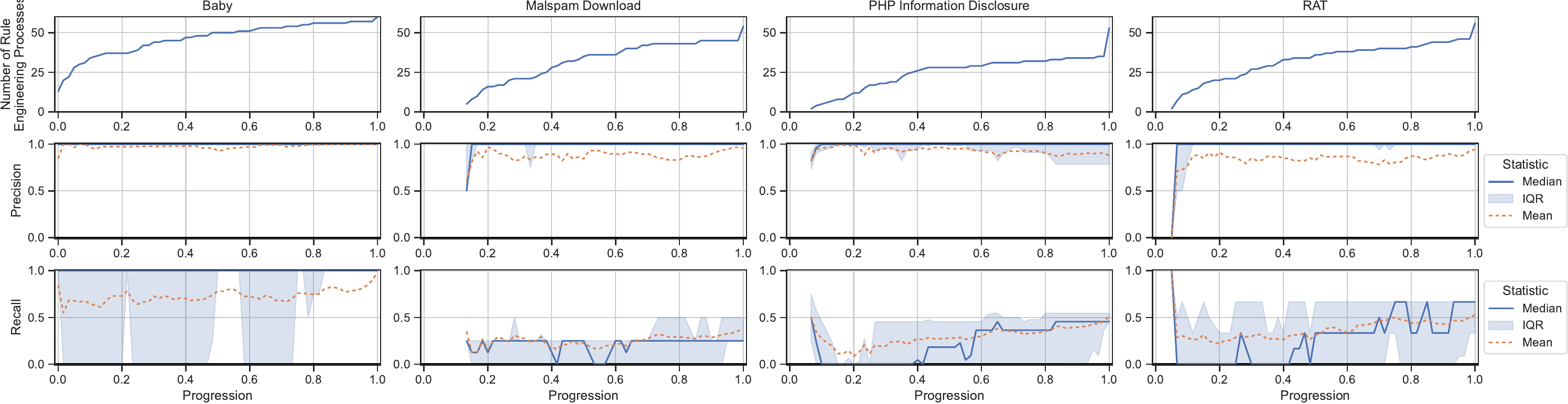}
    \centering
    \caption{Progression of participants for each main scenario showing the number of participants submitted a valid rule (top), and the Precision (middle) and Recall (bottom) on all tests of their most recent submission.}
    \label{fig:progression}
    \Description{Multi-panel progression plots for each main scenario. The top row shows the number of participants with a valid rule over the fraction of time spent. The middle row shows precision of the latest rule, and the bottom row shows recall. Precision is generally high early, while recall increases later.}
\end{figure*}

\cref{fig:progression} shows the participants' progression on the main scenarios, with the x-axis representing the fraction of time spent towards the final submission.
The top row shows how many participants have submitted a valid rule for that scenario. Note that many participants are initially not implementing a rule but instead investigating the behavior they should detect and considering how to engineer the rule(s). We identify this as \textbf{Phase~1} of the rule engineering process. We observe a sharp increase in participants who have submitted a valid rule near the end, suggesting these participants spent significant time thinking before implementing a rule.
Note that the distributions in the bottom two rows are computed over increasingly many samples over time as more participants have submitted a valid rule.
When inspecting the bottom two rows, it becomes evident that participants achieve good specificity early on as suggested by the high precision scores\footnote{\scriptsize In the case where a rule triggers zero alerts, a precision score of $1$ is assigned.}, whereas coverage is often lacking until the end of their progression as suggested by the low recall scores that only increase near the end. Coverage is also the characteristic of rules that participants end up improving the most as suggested by the trends. We identify the implementation of a rule that detects the relevant behavior as \textbf{Phase~2} of the rule engineering process.
We manually inspected several rules with simultaneous low coverage (recall $0.0$) and high specificity (precision $1.0$), to determine potential explanations for this peculiar performance. Some rules specify the wrong traffic direction in the rule header (i.e., IP address groups or port numbers). Others contain \textit{pcre} regular expressions that contain minor bugs causing them to not match any relevant traffic. Other rules have appropriately defined a characteristic to detect, but failed to specify where or how data to be matched should be preprocessed (i.e., not applying keywords like \textit{file.data} to prepare an HTTP body for processing).

\paragraph{Progression beyond passing visible tests}

\label{sec:progression_beyond_visible_tests}

\begin{figure*}[htbp]
    \includegraphics[width=1.0\textwidth]{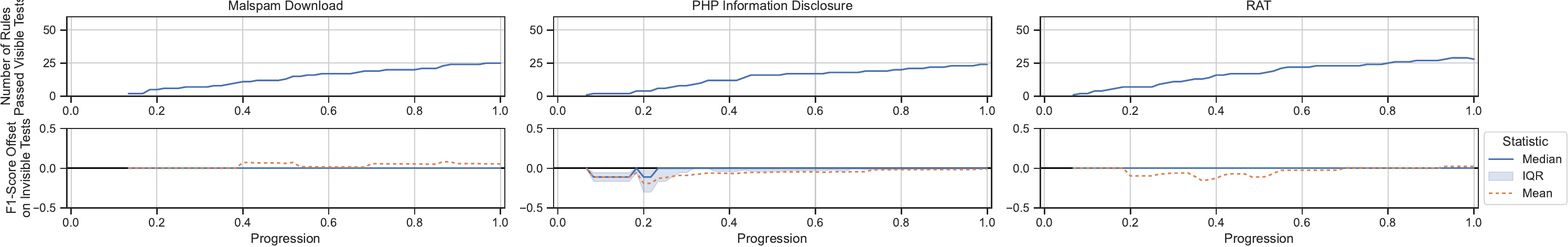}
    \centering
    \caption{Progression on main scenarios after passing the hidden tests showing how many participants pass the visible tests (top) and the offset in F1-score on the hidden tests of their most recent rule w.r.t. their first rule passing the visible tests (bottom).}
    \label{fig:progression-beyond-visible-tests}
    \Description{Multi-panel plots for the three main scenarios after participants first pass the visible tests. The top row shows the number of participants who continue submitting valid rules that pass the visible tests. The bottom row shows the change in hidden-test F1-score relative to each participant's first rule passing the visible tests.}
\end{figure*}

\looseness=-1
During the CTFs and while speaking with participants, we observe that participants struggle to improve the performance of rules on the hidden tests. We identify improving a rule beyond the available test data as \textbf{Phase~3} in the rule engineering process.
Interestingly, this observation relates to the impact of availability of test data (i.e., access to hidden tests) on the rule engineering process.
Verifying this observation using the data, we zoom in on the process participants follow after passing the visible tests (i.e., obtaining an F1-score of $1.0$ thereon). Concretely, we compare the performance on the hidden tests after having passed the visible tests to that when the visible tests are first passed.
\Cref{fig:progression-beyond-visible-tests} shows resulting comparisons, where the different columns cover the three main scenarios. The top row indicates how many participants have passed the visible tests and submitted subsequent valid rules also passing these tests.
The bottom row depicts the distribution of offsets in the F1-score on the hidden tests, i.e., the difference between the score of the most recent submission and that of the first submission to pass the visible tests.

\begin{figure}[htbp]
    \includegraphics[width=1.0\linewidth]{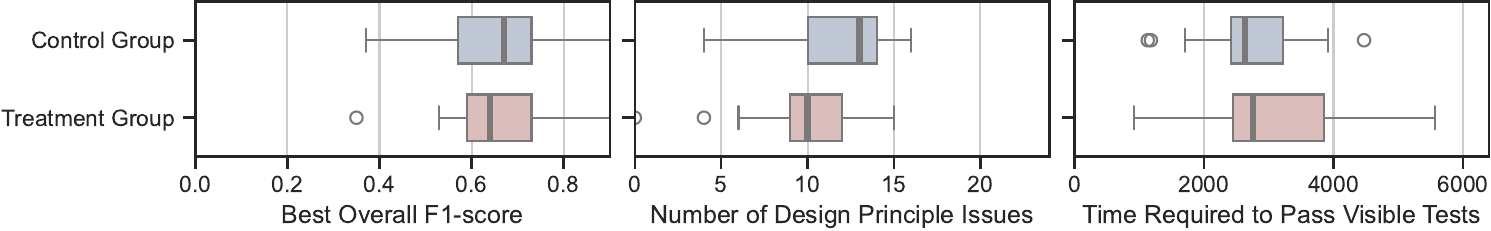}
    \centering
    \caption{Comparison of the control and treatment groups.
    }
    \label{fig:control-treatment-comparison}
    \Description{Three side-by-side box-plot comparisons of control and design-principles treatment groups. The panels show F1-score, number of design-principle issues, and time required to first pass the visible tests. The groups have similar F1-scores and completion times, while the treatment group has slightly fewer design-principle issues.}
\end{figure}

When inspecting the top row of \cref{fig:control-treatment-comparison} w.r.t. how many participants submitted a valid rule as in \cref{fig:progression}, we observe that a significant number of participants never proceeds to this phase since they submit no new valid rules passing the same visible tests after initially passing them.
The bottom row of \cref{fig:progression-beyond-visible-tests} suggests that most participants do not significantly or consistently improve rule performance (i.e., F1-score on all tests combined) beyond the already obtained performance.
Attempts of participants to generalize rule performance without knowledge on hidden tests can harm performance as indicated by the sharp decline in mean F1-score offset for the Information Disclosure scenario, which later averages out closer to zero as more participants pass the visible tests.

\paragraph{Rule Engineering Phases}
In \cref{sec:experiment_progression} and \cref{sec:progression_beyond_visible_tests} we identified three phases of the rule engineering process.
The phases can be summarized as: \textbf{\textit{(1)} plan, \textit{(2)} perform, \textit{(3)} perfect}, whereby rule engineers first plan the rule design, then perform the implementation, and finally perfect it to generalize.
\cref{fig:phases} (Appendix~\ref{app:rule_engineering_phases}) shows how many participants are in a certain phase.
The number of participants in Phase~1 initially remains constant and then quickly decreases, after which the decline stagnates. About half of the participants move on from Phase~1 within about $35\%$ of the time they spent on a scenario in total, suggesting that on average $35\%$ of their total time is spent thinking about how to design a signature for a scenario.
The number of participants in between Phases 1 and 2 remains overall low, suggesting that few participants struggle with the Suricata syntax.
The number of participants in Phase~2 gradually increases as people move on from Phase~1, and the number of participants in Phase~3 begins to increase shortly thereafter.
Importantly, we note that only about half of all participants eventually progress to Phase~3. We observe that most participants remaining in Phase~2 submit no new rule after having passed the visible tests. Several possible explanations exist for this phenomenon. Some participants may choose to quickly move on to the next scenario once they have built a rule that passes to visible tests, to obtain higher (visible) scores faster for the CTF. Others may be unsure how to improve rules and therefore not initiate such attempts. Other participants may not have time to attempt improvements for all scenarios.

\paragraph{Subject background and the effect of rule design principles}
\label{sec:experience_treatment}
To evaluate whether the background knowledge of a participant helps in engineering effective rules, we consider both their previous knowledge, as gauged with the questionnaire described in \cref{sec:experiment_design}, and their knowledge of rule design principles introduced in \cite{ruling-the-unruly} (i.e. whether they receive the training material).
To assess the effect of experience on the rule engineering process, we regress over scores for various experience categories on the questionnaire to predict three key performance metrics: \textit{(1)} the F1-score, \textit{(2)} the number of design principle issues detected by suricata-check\cite{suricata-check}, and \textit{(3)} the time required to pass the visible tests. The first two metrics relate to the quality of the engineered rule and the third metric pertains to the cost associated with the rule engineering process.
The regressions do not suggest prior experience has a significant effect on the performance of the $60$ participants with respect to any of the three metrics.
Therefore, experience appears to have negligible effects on performance of the rule engineering process and of resulting rules.
The regressions, together with $95\%$ confidence intervals, are presented in Appendix~\ref{app:experience_effects_regression}.

To assess how knowledge of the rule design principles impacts performance, we compare the metrics between the group that received the related instruction material and the one that did not.
\cref{fig:control-treatment-comparison} shows the performance of the two groups.
Subfigure \textit{(1)} shows that the participants in the treatment group have similar F1-scores. According to a Mann-Whitney U test, the difference is non-significant.
Subfigure \textit{(2)} shows that participants in the treatment group engineer rules that generally, albeit marginally, adhere better to the design principles ($p=0.04$) than those of users in the control group.
Subfigure \textit{(3)} shows that there are no clear statistical differences between the two groups in terms of time spent to first pass the visible tests.
From these data, it emerges that the treatment and control groups only minimally differ in adherence to design principles, which does not lead to significant differences in the performance of the rules. However, our experiment suggests that the adoption of design principles may lead to a trade-off with increased engineering time although we did not observe a clear statistical difference. This suggests more research is needed on how to train rule engineers and to balance the trade-off between increase in performance and time/resources needed to engineer rules. Due to the similar performance between the two groups, we disregard the distinction between groups for the remainder of the paper.

\paragraph{Uncovering rule engineering strategies}
\label{sec:rule_engineering_strategies}
Rule engineers follow different strategies to implement rules during Phase~2 and Phase~3. To identify different strategies for implementing rules, i.e., the order in which rule keywords are addressed during implementation, we use unsupervised learning to cluster similar strategies and then describe the clusters. 
First, rule implementation processes are modelled as variable-length sequences of categorical variables representing the steps taken in the process (i.e., addressed ).
Then, Sequence Graph Transform (SGT)~\cite{ranjan22-sequence-graph-transform-sgt} embeds these sequences into fixed-length numerical vectors. SGT leverages the probability that a categorical variable is followed by another categorical variable within some distance to encode the order of keywords addressed in each process.
Finally, HDBSCAN~\cite{mcinnes17-hdbscan} derives clusters, representing strategies, which we characterize to understand of how people engineer rules.

To model rule implementation processes as variable-length sequences of categorical variables, we automatically trigger whenever a participant checks or submits a new rule to detect which fields are added, modified, or removed since the previous check or submission.
Many rule fields are protocol-specific and scenarios use different protocols. Thus, we cannot use the fields directly to synthesize process steps across scenarios. 
To allow across-scenario analysis, in collaboration with two SOC experts, we define a mapping from the low-level Suricata fields to a high-level \textit{field type}.
Field types either \textit{(1)} specify the inspected \textit{traffic direction} (e.g., \texttt{src\_ip} in the header and \texttt{flow}), \textit{(2)} control \textit{payload matching} (e.g., \texttt{content}, and \texttt{depth}), \textit{(3)} deal with \textit{application layer} protocol aspects (e.g., \texttt{tls.cert\_issuer} and \texttt{base64\_decode}), \textit{(4)} add \textit{stateful} components (e.g., \texttt{threshold} and \texttt{flowbits}), or \textit{(5)} describe \textit{non-functional} aspects (e.g., \texttt{msg} and \texttt{fast\_pattern}). 
To create the sequences, we simply concatenate the field types of all changes in chronological order\footnote{\scriptsize If an update contains multiple changes we also apply alphabetical order.}. This resulted in $223$ sequences.

We use SGT to transform categorical variable-length sequences into $16$-dimensional numerical length-sensitive embeddings capturing long-term interdependencies between categorical variables.
We use HDBSCAN with Manhattan distance, which is suitabile for high dimensionalities~\cite{aggarwal01-distance-metrics-high-dimensionality}, to perform unsupervised clustering of embeddings and derive homogeneous clusters of engineering processes.
This process\footnote{\scriptsize Unless specified otherwise, default hyperparameters are used.} results in the five clusters/strategies shown in \cref{fig:clusters} (\cref{app:detailed_cluster_overviews}).
HDBSCAN rejected $152$ engineering processes, suggesting that many processes cannot be grouped into a homogeneous cluster, and there are potentially many possible strategies.
Cluster~$4$ is the largest cluster with $28$ processes and represents the most common strategy, whereas Cluster~$1$ is the smallest cluster with $7$ processes.
In Subfigure \textit{(1)} of \cref{fig:clusters}, Cluster~$4$ appears isolated from most other processes.
Subfigure \textit(2) shows no strong concentration of processes by scenario, suggesting specific scenarios do not necessarily prescribe following a certain strategy.
Only Clusters $1$ and $2$ do not contain processes from all scenarios.

\begin{figure*}[htbp]
    \centering
    \includegraphics[width=1.0\textwidth]{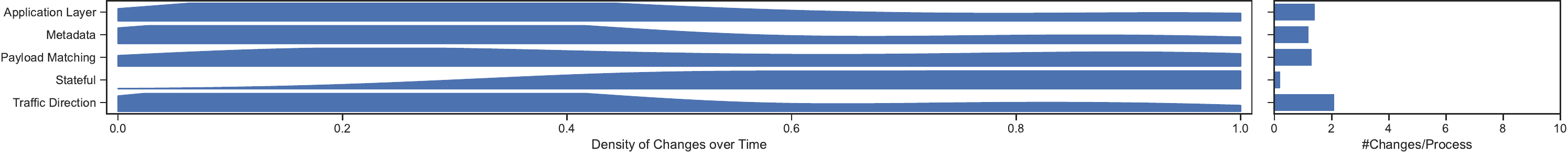}
    \caption{Summary of expert processes showing changes made over time (left), and number of changes per category (right).}
    \label{fig:expert-cluster-details}
    \Description{Summary of expert rule-engineering processes. The left plot shows when changes to rule-field categories occur over the normalized engineering process, and the right plot shows the number of changes in each category: non-functional, traffic direction, application layer, payload matching, and stateful fields.}
\end{figure*}

\cref{fig:expert-cluster-details} provides a density distribution of the changes applied to the submitted rules by the expert group. Additionally, it shows how many such changes are made.
Similar figures for all the other clusters discussed here are in Appendix~\ref{app:detailed_cluster_overviews}.
One commonality in strategies between the expert group and most of the clusters pertains to the order of changes to field types.
Most clusters including the expert group initially focus on fixing the traffic direction.
Application-layer and payload-matching changes follow, with approaches varying across clusters. This appears most applicable to Cluster~$4$, but also to Clusters $2$ and $3$, albeit with a more uniform approach, whereby changes are applied throughout.
Stateful aspects of rules (e.g., alert throttling) tend to be the last thing participants focus on for all clusters. A stark example is Cluster~$4$ wherein changes to stateful properties of the rule are solely made at the end.

Overall, the different clusters highlight very different strategies. 
Clusters $0$ and $1$ represent `one-shot' rule engineering strategies, whereby all changes are applied mostly at the end. Cluster~$2$ adopts a U-shaped approach whereby most changes are applied at the start and end, suggesting a more analytical approach towards the middle of the engineering process and a fine-tuning at the end. Cluster~$3$ applies changes throughout in a uniform fashion, without clear spikes but with relatively many submissions. This suggests an engineering process focused on a continuous testing-improving cycle, whereby engineers continuously test new changes and make improvements as they apply them. Cluster~$4$ strikes a middle ground between Cluster~$2$ (U-shaped process) and Clusters $0$/$1$ (most changes at the end). This suggests a relatively high prevalence of fine-tuning towards the end but with a more gradual rule engineering process.

Clusters also differ in terms of the rule features they implement. For example, Cluster~$1$ differs from Cluster~$0$ as it does not modify any stateful property of the rule. 
Cluster~$2$ differs from the other clusters in that no explicit payload matching keywords and no stateful keywords are used. Rules resulting from this strategy may instead rely on matching of IP addresses, which are specified as part of the traffic direction, which is the field type modified the most for Cluster~$2$.
Cluster~$3$ does not modify any stateful keywords and contains more overall modifications than other clusters.
Cluster~$4$ is most similar to the expert group, although the expert strategy makes modifications earlier, specifically for the stateful keywords.

The expert group first focuses on non-functional and traffic direction keywords, closely followed by application layer keywords, after which they turn to payload matching and finally on stateful keywords. The order in which experts address various keyword groups appears to be better separated, although similar to Cluster~$4$. Specifically, the distinction in order between payload matching and application layer keywords is more explicit for the expert strategy.
Note that experts consistently engineers rules with high F1-scores.

\begin{figure*}[htbp]
    \centering
    \includegraphics[width=1.0\textwidth]{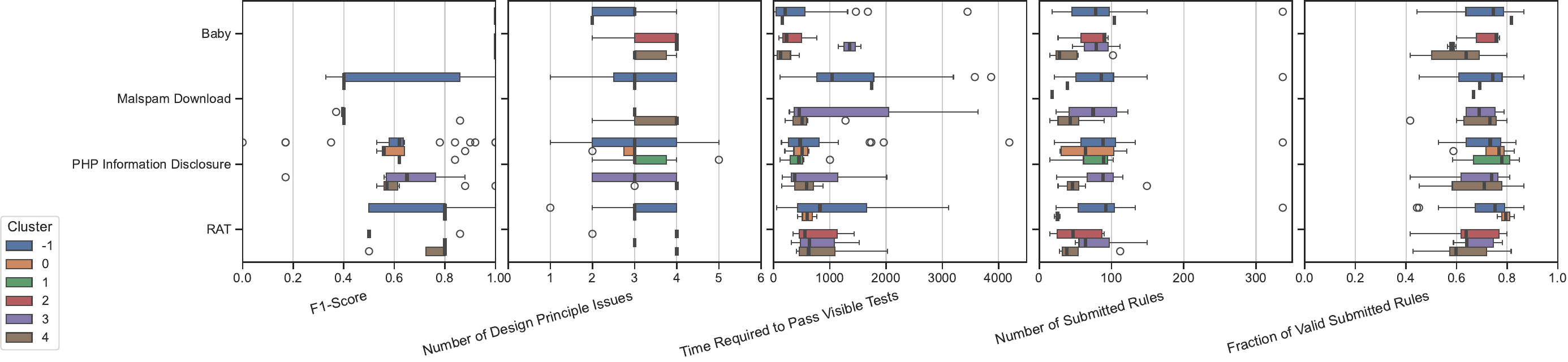}
    \caption{Comparison of the five clusters and the rejection cluster within each scenario.
    }
    \label{fig:clusters-comparison}
    \Description{Multi-panel comparison of five rule-engineering strategy clusters and rejected processes across the study scenarios. The panels compare F1-score, design-principle issues, time to pass visible tests, number of submissions, and fraction of valid submissions.}
\end{figure*}

\cref{fig:clusters-comparison} compares the effectiveness of the different strategies represented by the clusters.
Most performance measures obtained from the various clusters of engineering processes are similar for most scenarios, and where dissimilar no one strategy consistently under- or outperforms.
Subfigure \textit{(4)} suggests that processes from Cluster~$4$ have relatively few submitted rules in all scenarios, possibly indicating a well-thought-out process. Overall, we find no clear difference in outcomes across strategies, suggesting that forcing (e.g., through training or technological aids) a specific process upon rule engineers may lack tangible benefits in outcome. This suggests that rule engineering is both an engineering and a creative processs.

\paragraph{Fine-tuning behavior}
\label{sec:finetuning}
We further characterize rule updates in rule engineering processes by examining the edit distance using two metrics. The Levenshtein distance, relative to the rule length prior to the update, quantifies the portion of characters that was changed in an update\footnote{\scriptsize Note that addition of a single character (e.g., \textit{!}) may reverse the detection logic.}.
Secondly, multiset-Jacard similarity~\cite{jaccard-similarity} measures similarity between the rule design before and after the update in terms of the Suricata keywords used.
Unlike the Levenshtein distance, the Jaccard similarity insensitive to the Suricata syntax.

\begin{figure}[htbp]
{\centering%
\includegraphics[width=0.4\linewidth]{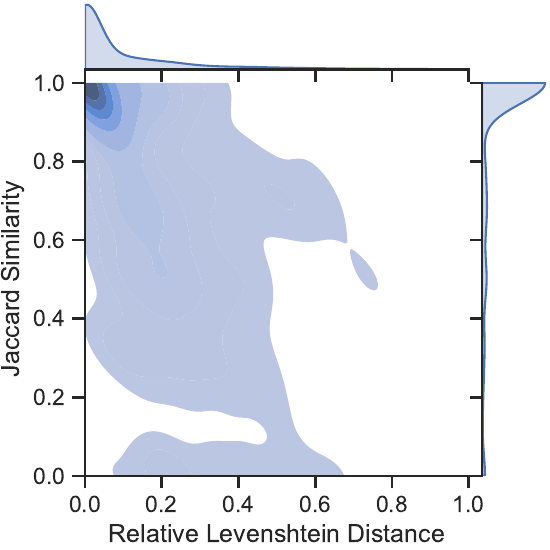}
\caption{Edit distances of rule updates.}
\label{fig:finetuning}
\Description{Joint distribution of rule-update edit distance measures. The horizontal axis shows relative Levenshtein distance, representing the fraction of characters changed, and the vertical axis shows Jaccard similarity of the Suricata keywords before and after an update. Concentration near low edit distance and high similarity indicates fine-tuning.}
}
\end{figure}

\cref{fig:finetuning} depicts the joint distribution of the relative Levenshtein distance to the Jaccard similarity. The figure suggests that the most updates only marginally change the rule as observed through the low relative Levenshtein distance and do not alter the rule design as indicated by the peak frequency of Jaccard similarities at $1.0$. These minimal rule updates reflect what we refer to as \textit{fine-tuning behavior}, meaning rule engineers primarily gradually refine rules over time, rather than pivoting to entirely different designs.

\section{Discussion}
\label{sec:discussion}

Although rule engineering is not trivial, as reflected by the low recall of detection rules (\cref{sec:experiment_progression}, the results of \cref{sec:experience_treatment} suggest that experience has no significant effects on the quality of the rule or the efficiency of the process by which that rule was derived.
This suggests that inexperienced rule engineers may be capable of effectively engineering  detection rules of sufficient quality, following a basic instruction and even with minimal background.

However, our results also emphasize the importance of test data availability to rule engineering, as participants appear unable to generalize rules after passing the visible tests.
In other words, it is difficult for rule engineers to develop signatures that generalize beyond the incidents and environments for which a rule was engineered. We find that increasing opportunities for rule engineers to test against more environments or incidents, or feedback loops as previously suggested~\cite{alert-alchemy} may promote rule quality and should be further explored in future research. Especially considering that many rules never receive updates~\cite{ruling-the-rules}, the initial engineering process is crucial for rule quality.
SOCs may benefit from employing methods to increase test data availability. For example, SOCs may integrate feedback loops to report FPs with test data to engineers to enhance the generalizability of rules.
Increased automation for rule engineering, including the collection of and testing on data, and possibly automatic generation of rules, is another opportunity.

Since results from \cref{sec:experience_treatment} suggest experience has no significant effect on rule engineering performance, rulesets established through community efforts as described in \cref{sec:rulesets} may contain rules of similar quality to commercial rulesets if the experience of rule engineers is the only differentiating factor.
Therefore, the idea of complementing internal rulesets with external (e.g., open) rulesets to increase coverage~\cite{alert-alchemy} may benefit SOCs.
Future research may further explore such differences, specifically w.r.t. the test data availability. Although rule-level quality may be similar, ruleset-level quality may differ as a result of how the ruleset is managed.

We identified three rule engineering phases and a common approach followed in the implementation, which is shared with the expert group.
The phases are discussed in \cref{sec:results} whereby rule engineers first plan the design of a rule, then perform the implementation, and finally perfect the rule.
We found several commonalities between all implementation processes (i.e., phases 2 and 3), suggesting the existence of a common strategy.
Rule engineers first focus on aspects surrounding detection characteristics, such as the circumstance in which a characteristic can be observed by implementing the direction of the network traffic. Subsequently, they move on to implementing the detection of the characteristic(s) itself using application layer and payload matching keywords.  Cluster~$4$, which follows this approach and is the largest cluster, shows a high similarity to the expert group in terms of the strategy followed.
SOCs may organize internal processes around these phases and the common strategy to streamline workflows.
Moreover, we found that rule engineers focus updates on fine-tuning rules as opposed to pivoting to alternative rule designs. On the other hand, our results suggest that rule engineering is as also a creative process, following different strategies, focusing on different aspects of rule engineering, which appear to remain largely equivalent in terms of performance outcomes. This suggests that strict training or engineering processes, for example enforced through the use of supporting technology, may lead to negligible or no benefit in terms of rule quality and may boil down to solely a standardization effort. Rule engineering training and support could therefore focus on rule completeness rather than the engineering processes, for instance, by ensuring all relevant aspects like design principles are considered. Further research is needed in this direction.

\noindent\textit{Limitations.}
Due to the high similarity between Suricata and Snort, we expect our findings to be directly applicable to Snort rules. Although Suricata and Snort function as NIDS, other rule languages such as YARA~\cite{yara} and Sigma~\cite{sigma} function in different operational contexts and use different syntax. While these differences may limit the transferability of findings, the insights provided in this work retain relevance since the starting point (i.e., benign and malicious tests) and the goal (i.e., detecting malicious behavior) remain similar. We believe that \textsc{SuriCap} can be extended or used as a blueprint to study related different rule languages or different operational contexts, such as those offered by Zeek~\cite{Zeek}, YARA, or Sigma.

Although the subjects of our study are primarily novices in detection engineering, which are not most representative of typical SOC rule engineers, we believe that our findings present meaningful insights that are applicable to SOC operations since experience does not significantly affect rule engineering processes according to our findings in \cref{sec:experience_treatment}. Moreover, we compared the rule engineering processes of several SOC experts in \cref{sec:rule_engineering_strategies} and found them to be similar to those of most processes in our study. Previous research also studied novices to better understand security-related tasks~\cite{decomperson,10917883}. Furthermore, we remark that many rules (specifically community-sourced) are not necessarily written by experienced rule engineers as discussed in \cref{sec:rulesets} and that typical SOC hires have experience-levels similar to our participants~\cite{kathryn-22}.

While our work addresses the rule engineering aspect of detection engineering, the processes detection engineers follow to select detection characteristics (i.e., Phase~1) is not captured by \textsc{SuriCap}. Although we acknowledge that characteristics chosen significantly impact coverage and specificity~\cite{ruling-the-rules}, studying how detection engineers select characteristics would be intrusive (i.e., employing screen captures) or require significant efforts in developing data collection methods for interactions with applications like Wireshark, \textsc{tcpdump}, or web browsers. Although other research methods could be applied, they complicate data collection at-scale, which is preferred due to the large variety in rule engineering strategies as discussed in \cref{sec:rule_engineering_strategies}.
Nevertheless, Phase~1 and may be an interesting topic for future research given that rule engineers rarely pivot to alternative rule designs as discussed in \cref{sec:finetuning}.

\looseness=-1
While the presented data analysis is limited, we shed light on the effects on rule engineering processes of prior experience, training, and test data availability.
Enabling future research, we release the PCAPs used to evaluate rules, the platform used to organize CTF-style workshops, the collected datasets, and the code used to conduct the presented analysis.
We believe the integrated development environment and continuous feedback provided by \textsc{SuriCap} may be beneficial for training or otherwise improve rule engineering processes similar to what is typically offered by cyber ranges \cite{vielberth2021digital, glas2022visual}. To support such endeavors, we release \textsc{SuriCap}\footnote{\scriptsize \textit{\url{https://github.com/Koen1999/suritrain}}} to the public.

\section{Conclusion}
\label{sec:conclusion}

This work presents \textsc{SuriCap} as a novel platform to study NIDS rule engineering and used it to collect data, from which we derive the first academic characterization of detection engineering processes. We identified rule engineering phases, a common implementation strategy, and shed light on the importance of available test data and fine-tuning behavior, to benefit researchers and SOCs.

\begin{acks}
The authors thank colleagues at other institutions who helped organize the workshops, and everyone who participated in them.
This paper was edited for grammar using LanguageTool and Writefull.
This publication is part of the CATRIN and INTERSECT projects (with numbers {\small NWA.1215.18.003} and {\small NWA.1160.18.301}), which is (partly) financed by the Dutch Research Council (NWO).
\end{acks}

\bibliographystyle{ACM-Reference-Format}
\bibliography{main}

\appendix

\section{Instruction contents}
\label{app:instruction}

The preparatory instruction first covers the experiment design (including the existence of hidden tests), and the general experience the participants will undergo.
Practical aspects of Wireshark~\cite{Wireshark}, a tool to inspect network traffic captures, are covered by discussing common features such as protocol recognition, display filters, and the following of streams.
Subsequently, the concept of intrusion detection is generally introduced and Suricata is covered in-depth by explaining the rule syntax and highlighting common functional keywords including \texttt{content}, \texttt{file.data}, \texttt{flow}, \texttt{flowbits}, and \texttt{threshold} as well as transformations, modifiers, and protocol-specific keywords for DNS, HTTP, and TLS. The existence of other options is emphasized with reference to the official Suricata documentation~\cite{Suricata-docs}, which is also accessible through the platform UI.
We exemplify how a simple rule containing all keywords can be written.
The instruction also explains the various scenarios that users tackle during the activity, highlighting why one would want to detect such scenarios and possibly relevant information without prescribing detection characteristics.
The general instruction ends with practical details on the employed platform and the leaderboard scoring method, as well as a live demonstration of the platform.
The general instruction takes approximately one hour.

\section{Intake Questionnaire}
\label{app:questionnaires}
\label{sec:experience_assessment}
To grade the intake questionnaire, we group questions of the same topic.
Furthermore, we distinguish two types of multiple choice questions: \textit{(1) single-answer questions} to which only a single answer can be selected, and \textit{(1) multiple-answer questions} to which multiple answers may be selected and that have zero or more valid answers. For single-answer questions, participants receive a score of one if the correct answer is selected and zero otherwise. For multiple-answer questions, the score equals the number of selected correct answers minus the number of incorrect answers, with a minimum of zero.
The contents of the intake questionnaire are found hereafter.\\

\label{app:questionnaire}
{\small
\textbf{Questions 1--4} collected the participant's contact address, leaderboard username, confirmation of having the legal age to consent, and the informed consent.\\
\noindent\textbf{Question 5:} How did you first learn about the CTF activity?
\begin{enumerate}[label=(\alph*)]
  \item \textsc{Promotion channel 1 (e.g., course)}
  \item \textsc{Promotion channel 2 (e.g., CTF association)}
  \item $\hdots$
\end{enumerate}
\noindent\textbf{Question 6:} What is your gender?\\
\begin{enumerate*}[label=(\alph*)]
  \item Male
  \item Female
  \item Other
  \item I prefer not to disclose
\end{enumerate*}\\
\noindent\textbf{Question 7:} What is the highest level of education that you have completed?
\begin{enumerate}[label=(\alph*)]
  \item Primary education (ISCED 1)
  \item Lower secondary education (ISCED 2)
  \item Upper secondary education (ISCED 3)
  \item Post-secondary non-tertiary education or Short-cycle tertiary education (ISCED 4 or 5)
  \item Bachelor's or equivalent (ISCED 6)
  \item Master's or equivalent (ISCED 7)
  \item Doctorate or equivalent (ISCED 8)
\end{enumerate}
\noindent\textbf{Question 8:} What is your level of expertise regarding Capture the Flags (or similar activities)?
\begin{enumerate}[label=(\alph*)]
  \item I have never heard of a Capture the Flag.
  \item I know what a Capture the Flag is but have never participated in one.
  \item I have previously participated in a Capture the Flag.
  \item I have participated in multiple Capture the Flags.
  \item I have previously organized a Capture the Flag.
\end{enumerate}
\noindent\textbf{Question 9:} Do you have any other relevant experiences (such as a job related to Security Operations or high expertise in another area you deem relevant) that may be relevant for the research? If so, describe the topic and the relevant experience below.\\
\noindent\textbf{Question 10:} How does the TCP protocol ensure that the packets are delivered in the correct order?
\begin{enumerate}[label=(\alph*)]
  \item Through the use of SYN and ACK flags.
  \item Through the encryption and decryption of data packets.
  \item Through the use of sequence and acknowledgment numbers.
  \item Through the use of checksums.
  \item I do not know.
\end{enumerate}
\noindent\textbf{Question 11:} What types of information can be exchanged during the initiation of a TLS session?
\begin{itemize}[label=\textbf{$\square$}]
  \item Plain-text HTTP requests
  \item Server and client Hello messages
  \item Encrypted handshake messages
  \item Certificate
  \item I do not know.
\end{itemize}
\noindent\textbf{Question 12:} Which of the following statements regarding network scanning are true?
\begin{itemize}[label=\textbf{$\square$}]
  \item Lack of response (SYN,ACK) to a TCP SYN implies a port is closed.
  \item If a vulnerability scanner detects a vulnerability, the scanned device has an exploitable vulnerability.
  \item Nmap can only perform UDP/TCP scans.
  \item If you want to know which ports may be open on a public-facing IP address, you must scan it.
  \item Network scanning may be used to inspect application layer features such as used software and versions.
  \item I do not know.
\end{itemize}
\noindent\textbf{Question 13:} Which of the following statements regarding Cross-Site Scripting (XSS) are true?
\begin{itemize}[label=\textbf{$\square$}]
  \item During a successful XSS attack malicious code is executed on a server distributing content.
  \item XSS attacks may be prevented through input sanitization.
  \item XSS attacks can be used to steal credentials.
  \item During a successful XSS attack the attacker may execute malicious code within the web browsers of other users visiting a website.
  \item XSS attacks are considered to be a subset of SQL injection attacks.
  \item I do not know.
\end{itemize}
\noindent\textbf{Question 14:} Which of the following statements regarding Persistence are true?
\begin{itemize}[label=\textbf{$\square$}]
  \item Persistence always implies that following a malware infection, malware will remain present after reinstallation of the operating system.
  \item Persistence must be obtained in order for malware to accomplish its goal.
  \item In order to obtain persistence, malware must add an executable file to the startup folder on Windows.
  \item Metasploit is a tool that may be used to obtain persistence depending on the targeted system.
  \item Persistence always has associated network traffic.
  \item I do not know.
\end{itemize}
\noindent\textbf{Question 15:} What is/are (a) feature(s) offered by Wireshark?
\begin{itemize}[label=\textbf{$\square$}]
  \item Filtering of packets
  \item Decoding of transmitted data
  \item Blocking network traffic
  \item Decrypting all encrypted network traffic
  \item Producing network I/O statistics
  \item Inspecting raw bytes transmitted in streams
  \item I do not know.
\end{itemize}
\noindent\textbf{Question 16:} What is the purpose of the ip.addr==192.168.178.1/24 filter in Wireshark?
\begin{enumerate}[label=(\alph*)]
  \item To only show network packets originating from the subnet 192.168.178.1/24 
  \item To only show network packets sent to the subnet 192.168.178.1/24
  \item To only show network packets originating from or sent to the subnet 192.168.178.1/24
  \item To only show network packets originating from and sent to the subnet 192.168.178.1/24
  \item It is an invalid filter.
  \item I do not know.
\end{enumerate}
\noindent\textbf{Question 17:} This question is just to check if you are actually reading questions carefully. It is an attention check. If you read this, please click the third option.\\
\begin{enumerate*}[label=(\alph*)]
  \item HTTP
  \item DNS
  \item SSH
  \item TLS
  \item UDP
  \item TCP
  \item I do not know.
\end{enumerate*}\\
\noindent\textbf{Question 18:} Which of the following statements regarding different intrusion detection paradigms are true?
\begin{itemize}[label=\textbf{$\square$}]
  \item Signature-based intrusion detection methods always rely on Atomic Indicators of Compromise (IOCs).
  \item Anomaly-based intrusion detection methods always rely on machine learning.
  \item Signature-based intrusion detection methods primarily rely on knowledge of malicious behaviors.
  \item Anomaly-based intrusion detection methods primarily rely on knowledge of benign behaviors.
  \item Anomaly-based intrusion detection methods are a subset of signature-based intrusion detection methods.
  \item I do not know.
\end{itemize}
\noindent\textbf{Question 19:} What are built-in functionalities offered by Suricata?
\begin{itemize}[label=\textbf{$\square$}]
  \item Matching bytes at specific locations
  \item Stateful detection across different flows in which the same IP address is involved
  \item Matching specific traffic directions
  \item Decoding of certain HTTP buffers
  \item Stateful detection within the same flow
  \item Matching using a remote API
  \item Matching using regular expressions
  \item I do not know.
\end{itemize}
\noindent\textbf{Question 20:} Which of the following buffers would be matched by the following sequence of Suricata options:
\noindent\textsc{content:"FOO"; content:"bar"; depth:5;}\\
\begin{itemize*}[label=\textbf{$\square$}]
  \item FOObar
  \item barFOO
  \item foobar
  \item barfoo
  \item I do not know.
\end{itemize*}\\
\noindent\textbf{Question 21:} Read the following Suricata documentation and select the factually correct statements one can derive from the given sentence.\\
\noindent\textsc{[Suricata documentation on \textit{startswith} and \textit{dotprefix}]}\\
\noindent Now consider a typical DNS request to \textit{google.com} Which of the following rules are valid rules and will match this request?
\begin{itemize}[label=\textbf{$\square$}]
  \item alert dns any any $\rightarrow$ any 53 (msg:"DNS Request to google.com"; dns.query; content:"google.com"; startswith; sid:1;)
  \item alert dns any any $\rightarrow$ any 53 (msg:"DNS Request to google.com"; dns.query; content:".google.com"; startswith; sid:1;)
  \item alert dns any any $\rightarrow$ any 53 (msg:"DNS Request to google.com"; dns.query; dotprefix; content:"google.com"; startswith; sid:1;)
  \item alert dns any any $\rightarrow$ any 53 (msg:"DNS Request to google.com"; dns.query; dotprefix; content:".google.com"; startswith; sid:1;)
  \item alert dns any any $\rightarrow$ any 53 (msg:"DNS Request to google.com"; dns.query; content:"google.com"; startswith; dotprefix; sid:1;)
  \item alert dns any any $\rightarrow$ any 53 (msg:"DNS Request to google.com"; dns.query; content:".google.com"; startswith; dotprefix; sid:1;)
  \item I do not know.
\end{itemize}
}

\section{Additional scenarios}
\label{app:additional_scenarios}
\paragraph{Baby}
To familiarize participants with Suricata and the CTF platform, and similarly to \cite{decomperson}, participants first encounter a \textit{Baby} `warm-up' scenario. This scenario requires participants to detect a DNS request to \textsc{detectme.foobar}. We manually collected a PCAP for this scenario using \textsc{nslookup}, which resulted in one \textsc{A} and one \textsc{AAAA} query. We instruct the participants to engineer a rule that detects DNS requests to this domain. Whereas simple, it allows participants to familiarize with the interface and provides the opportunity for a quick practical refresher on Suricata syntax.

\paragraph{Malspam Download}
This scenario covers the retrieval of a webpage that initiates a download for a malicious binary from the \textit{Guildma/Astaroth} campaign. We leverage four PCAPs containing HTTP traffic following the clicking through on malspam emails from this campaign, collected over a time span of over two years. We present participants with the oldest capture\footnote{\scriptsize \url{https://www.malware-traffic-analysis.net/2021/04/12/index.html}} for the visible tests. We used the three most recent captures\footnote{\scriptsize \url{https://www.malware-traffic-analysis.net/2021/07/02/index.html}}\footnote{\scriptsize \url{https://www.malware-traffic-analysis.net/2022/09/21/index.html}}\footnote{\scriptsize \url{https://www.malware-traffic-analysis.net/2023/12/11/index.html}} for invisible tests to assess the generalizability of rules to more recent incidents of the same campaign. Although all four captures include similar mechanisms to initiate a download, the most recent PCAP contains double base64-encoded HTML to initiate a download, indicating adversaries have become increasingly evasive over time. The first three PCAPs exhibit very similar behavior although specifics such as domain names, payloads, and some irrelavant data vary.

\paragraph{RAT}
\looseness=-1The \textit{RAT} scenario covers the outbound C2 communications of a Remote Access Toolkit (RAT) of the AsyncRAT family. These communications occur over TLS, making it more challenging than the other scenarios. Nevertheless, the communications preceding the encrypted TLS traffic offer various detection opportunities, amongst others due to the use of unusual certificates. We leverage two network traffic captures from 2024\footnote{\scriptsize \url{https://www.malware-traffic-analysis.net/2024/01/09/index.html}}\footnote{\scriptsize \url{https://www.malware-traffic-analysis.net/2024/03/14/index.html}} and an older capture of the AsyncRAT variant that we had previously collected in 2022 for educational purposes.
The most recent capture contains traffic from DcRat, which is a variant of AsyncRAT~\cite{asyncrat-branches}, amongst others varying certificate details.
Similarly to the other scenarios, we use the oldest available PCAP as visible test for the participant, and the more recent PCAPs as invisible tests to assess the generalizability of rules to more recent incidents corresponding to the same malware family.

\section{Experience effects regression}
\label{app:experience_effects_regression}
\begin{figure}[htbp]
\vspace{-0.05in}
    \centering
    \includegraphics[width=1.0\linewidth]{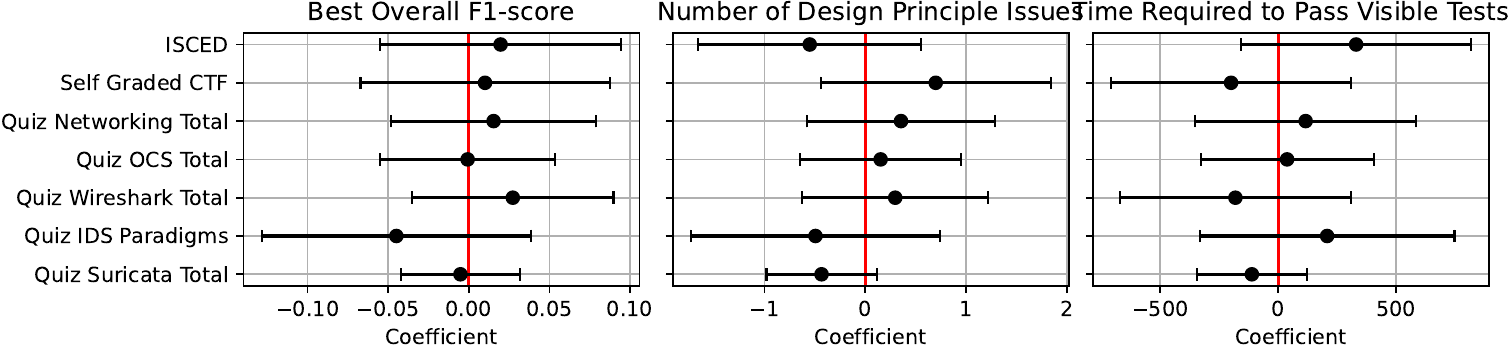}
\vspace{-0.15in}
    \caption{Three plots depicting the relation between the experience and their performance across all scenarios, each showing the coefficient together with confidence intervals.
    }
    \label{fig:experience-effects-new}
    \Description{Three coefficient plots showing the relationship between participant experience and performance across scenarios. Each point estimates an experience coefficient from a robust linear model, with a confidence interval, for F1-score, number of design-principle issues, or time to pass the visible tests.}
\vspace{-0.10in}
\end{figure}

\cref{fig:experience-effects-new} shows the regression\footnote{\scriptsize All VIF scores are below $2$.} using a Robust Linear Model ($p=0.05$) over the participants' experience to predict their performance as described and discussed in \cref{sec:experience_treatment}.

\section{Progression through rule engineering phases}
\label{app:rule_engineering_phases}
\begin{figure*}[h]
    \includegraphics[width=1.0\textwidth]{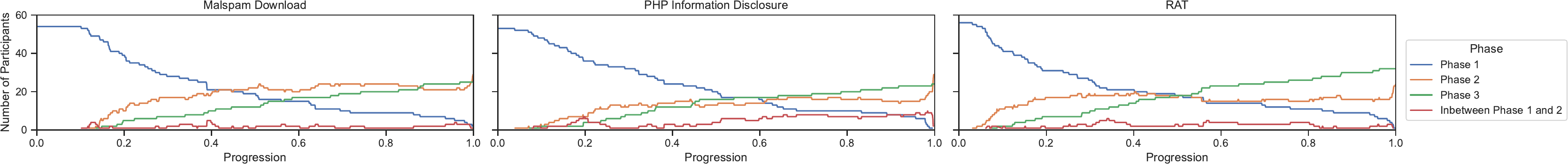}
    \centering
    \caption{Participants per rule engineering phase over time for each main scenario.}
    \label{fig:phases}
    \Description{Multi-panel plots showing the number of participants in each rule-engineering phase over normalized time for each main scenario. The phases represent planning, implementing a valid rule, and perfecting the rule beyond the visible tests.}
\end{figure*}

\cref{fig:phases} shows the number of participants in a certain phase.

\section{Detailed cluster overviews}
\label{app:detailed_cluster_overviews}
\begin{figure}[H]
    \includegraphics[width=1.0\linewidth]{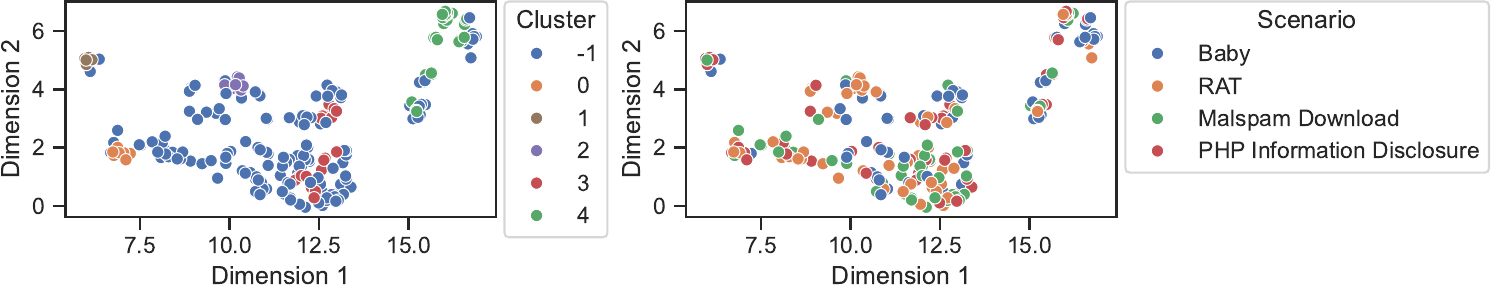}
    \centering
    \caption{Clusters of rule engineering processes where, the left highlights the cluster a process belongs to, and the right highlights the scenario a process was for.}
    \label{fig:clusters}
    \Description{Two-dimensional embedding of rule-engineering processes. The left plot colors each process by its assigned cluster or rejection status, while the right plot colors processes by the scenario in which they occurred.}
\end{figure}

\begin{figure*}[h]
    \centering
    \begin{subfigure}[b]{1.0\textwidth}
        \centering
        \includegraphics[width=1.0\textwidth]{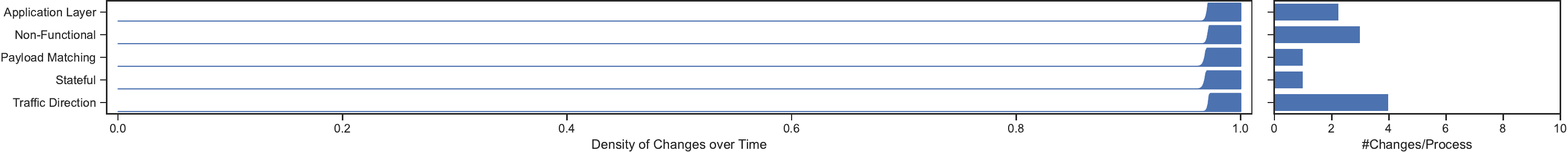}
        \vspace{-0.15in}
        \caption{Cluster 0}
    \end{subfigure}
    \begin{subfigure}[b]{1.0\textwidth}
        \centering
        \includegraphics[width=1.0\textwidth]{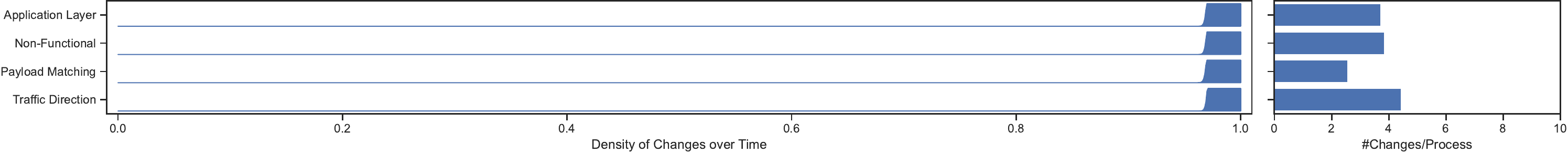}
        \vspace{-0.15in}
        \caption{Cluster 1}
    \end{subfigure}
    \begin{subfigure}[b]{1.0\textwidth}
        \centering
        \includegraphics[width=1.0\textwidth]{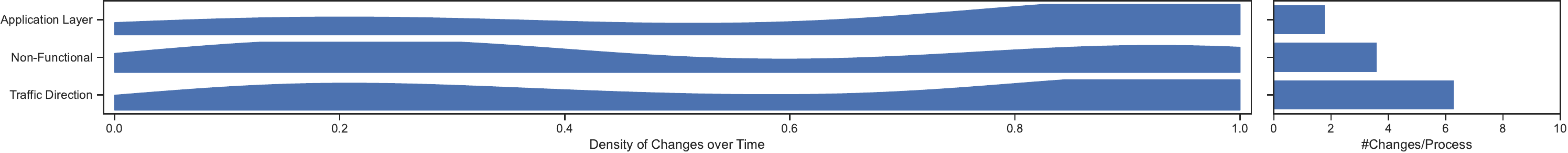}
        \vspace{-0.15in}
        \caption{Cluster 2}
    \end{subfigure}
    \begin{subfigure}[b]{1.0\textwidth}
        \centering
        \includegraphics[width=1.0\textwidth]{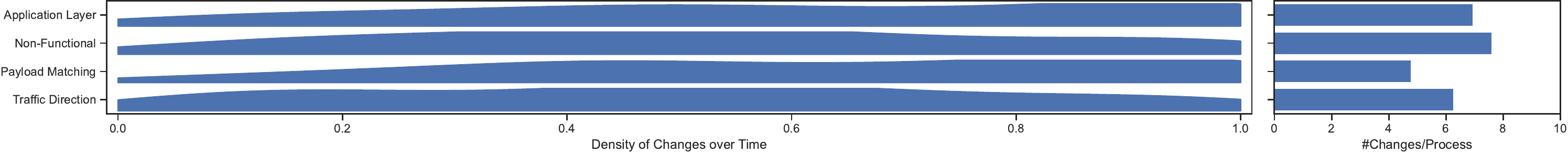}
        \vspace{-0.15in}
        \caption{Cluster 3}
    \end{subfigure}
    \begin{subfigure}[b]{1.0\textwidth}
        \centering
        \includegraphics[width=1.0\textwidth]{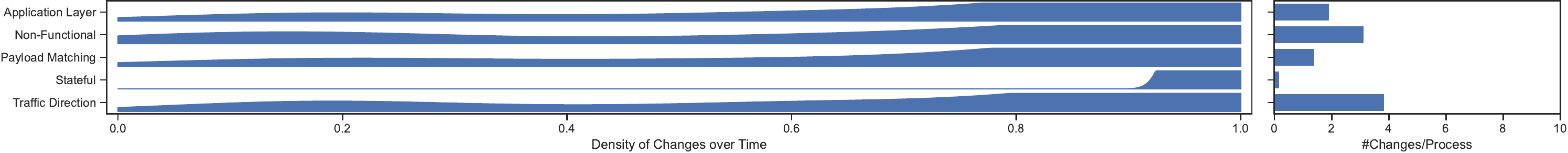}
        \vspace{-0.15in}
        \caption{Cluster 4}
    \end{subfigure}
    \begin{subfigure}[b]{1.0\textwidth}
        \centering
        \includegraphics[width=1.0\textwidth]{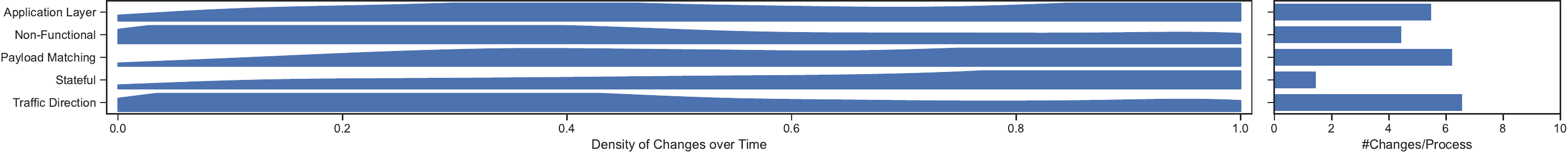}
        \vspace{-0.15in}
        \caption{Rejected engineering processes}
    \end{subfigure}
    \caption{Detailed overview of the five clusters of engineering processes and the group of rejected engineering processes, depicting within each subfigure, the changes made over time (left), and the number of changes per category (right).}
    \label{fig:clusters-details}
    \Description{Detailed summaries for clusters 0 through 4 and rejected engineering processes. Each subfigure contains a left plot showing changes to rule-field categories over the normalized process and a right plot showing the number of changes in each category.}
\end{figure*}

The five clusters/strategies of rule engineering processes resulting from and described in \cref{sec:rule_engineering_strategies} are shown in \cref{fig:clusters}\footnote{\scriptsize UMAP~\cite{mcinnes20-umap} is used with the Manhattan distance to reduce to $2$-dimensions.}.
\cref{fig:clusters-details} contains the detailed overview of the clusters described in \cref{sec:rule_engineering_strategies} similar to how the expert cluster is depicted in \cref{fig:expert-cluster-details}

\end{document}